\documentstyle[12pt,aaspp4]{article}

\begin{document}

%
\def\ltsim{\raisebox{-.5ex}{$\;\stackrel{<}{\sim}\;$}}
\def\gtsim{\raisebox{-.5ex}{$\;\stackrel{>}{\sim}\;$}}
\def\hi{H\,{\sc i}}
\def\hii{H\,{\sc ii}}
\def\hei{He\,{\sc i}}
\def\heii{He\,{\sc ii}}
\def\ciii{C\,{\sc iii}}
\def\civ{C\,{\sc iv}}
\def\nii{N\,{\sc ii}}
\def\oi{O\,{\sc i}}
\def\oii{O\,{\sc ii}}
\def\oiii{O\,{\sc iii}}
\def\ovi{O\,{\sc vi}}
\def\neiii{Ne\,{\sc iii}}
\def\nev{Ne\,{\sc v}}
\def\nevi{Ne\,{\sc vi}}
\def\navi{Na\,{\sc vi}}
\def\mgii{Mg\,{\sc ii}}
\def\mgv{Mg\,{\sc v}}
\def\mgvii{Mg\,{\sc vii}}
\def\mgviii{Mg\,{\sc viii}}
\def\alv{Al\,{\sc v}}
\def\alvi{Al\,{\sc vi}}
\def\alviii{Al\,{\sc viii}}
\def\sii{S\,{\sc ii}}
\def\siii{S\,{\sc iii}}
\def\sivi{Si\,{\sc vi}}
\def\sivii{Si\,{\sc vii}}
\def\siix{Si\,{\sc ix}}
\def\six{Si\,{\sc x}}
\def\sviii{S\,{\sc viii}}
\def\suix{S\,{\sc ix}}
\def\suxi{S\,{\sc xi}}
\def\ariii{Ar\,{\sc iii}}
\def\arvi{Ar\,{\sc vi}}
\def\arx{Ar\,{\sc x}}
\def\arxi{Ar\,{\sc xi}}
\def\caii{Ca\,{\sc ii}}
\def\caviii{Ca\,{\sc viii}}
\def\feii{Fe\,{\sc ii}}
\def\fevii{Fe\,{\sc vii}}
\def\fex{Fe\,{\sc x}}
\def\fexi{Fe\,{\sc xi}}
\def\fexiv{Fe\,{\sc xiv}}\def\n{\footnotemark}
\def\o{\o}
%

\title{Physical Conditions of the Coronal \\
Line Region in Seyfert Galaxies}

\author{Jason W. Ferguson, Kirk T. Korista, Gary J. Ferland}
\affil{Department of Physics \& Astronomy, University of Kentucky,
Lexington, KY 40506}

\begin{abstract}
The launch of the Infrared Space Observatory and new atomic data have
opened a window to the study of high ionization gas in active
galactic nuclei (AGN).  We present the results of a large number of
photoionization simulations of the ``coronal line'' region in AGN,
employing new atomic data from the Opacity and Iron Projects.  Our grid
of line emission spans 8 orders of magnitude in gas density and 14
orders of magnitude in ionizing flux in an effort to identify the
optimal conditions in which these lines form.  We show that coronal
lines form at distances from just outside the broad line region to
$\sim 400L_{43.5}^{1/2}$ pc, in gas with ionization parameter $-2.0
\ltsim \log U(H) \ltsim 0.75$, corresponding to gas densities of 10$^2$
to 10$^{8.5}$ cm$^{-3}$, with electron temperatures $\sim$ 12,000~K --
150,000~K.  A large range of distances from the central source
implies significant line width variation among the coronal lines.  We
identify several line ratios which could be used to measure relative
abundances, and we use these to show that the coronal line gas is
likely to be dust free.
\end{abstract}



\section{Introduction}
Given their high ionization potentials ($\chi$ $>$ 100 eV) the presence
of highly ionized optical forbidden lines, such as [\fevii\/]
$\lambda$6087, [\fex\/] $\lambda$6375, [\fexi\/]
$\lambda$7892, and [\fexiv\/] $\lambda$5303, in the spectra of
Seyfert galaxies points to very energetic processes at work in these
active galactic nuclei (AGN) (Oke \& Sargent 1968; Souffrin 1968;
Grandi 1978; Penston et al.\ 1984). If mechanically shocked and
collisionally ionized, this gas is as hot as our Sun's corona,
10$^6$~K, thus the origin of their name, ``coronal lines''. On the
other hand, if photoionized by the hard ionizing continuum of the AGN,
the coronal emission line gas is expected to be only a few to several
tens of thousands of degrees. (Nussbaumer \& Osterbrock 1970; Grandi
1978; Korista \& Ferland 1989; Oliva et al.\ 1994; Pier \& Voit 1995; 
Oliva 1996).

Recent observations of the optical coronal line profiles show that they
have FWHM broader than those of lower ionization forbidden lines, such
as [\oiii\/] $\lambda$5007 (DeRobertis \& Osterbrock 1984, 1986;
Appenzeller \& \"{O}streicher; Appenzeller \& Wagner 1991; Veilleux
1991).  This and the fact that the coronal emission line critical
densities are also larger ($10^7 - 10^{10}$~cm$^{-3}$) has led to
speculation that these lines form in a region intermediate between the
classical narrow line region and the broad line region.

New ground-based infrared observations have focused on the strongest
lines visible through infrared windows in the earth's atmosphere,
namely [\mgviii] 3.03$\mu$m, [\sivi\/] 1.96$\mu$m, [\sivii\/]
2.48$\mu$m, [\siix\/] 3.94$\mu$m, [\six\/] 1.43$\mu$m, [\suix\/]
1.25$\mu$m, and [\caviii\/] 2.32$\mu$m, (see Oliva \& Moorwood 1990;
Spinoglio \& Malkan 1992; Voit 1992; Oliva et al.\ 1994; Giannuzzo,
Rieke \& Rieke 1995; Marconi et al.\ 1996; Thompson 1996).  However,
the new Infrared Space Observatory (ISO) is now producing high quality
infrared spectra of Seyfert galaxies over a broad wavelength range
(Kessler 1996; Moorwood et~al.\ 1996).

With the advent of modern infrared spectroscopy and new atomic data
computations  from the ``Opacity Project'' (Seaton et al.\ 1992) and
the ``Iron Project'' (Hummer et al.\ 1993), the opportunity to
understand the origin and nature of this high ionization emission is
upon us. To this end we present the results of a large grid of
photoionization calculations in which we illustrate graphically those
parameters, radius from the central ionizing source and gas density,
which produce 24 coronal lines most efficiently. These will be
important tools for the quantitative spectroscopist interested in
understanding the physical conditions in which coronal lines arise.

\section{Model calculations}

\subsection{New atomic data base for the coronal lines} 
Recent years have seen the emergence of accurate computations of
collision strengths, lead by the efforts of the ``Iron Project''.  We
present model calculations of coronal emission lines using the data
from Lennon \& Burke (1994; [\nev\/], [\mgvii\/], [\alviii\/],
[\siix\/], [\suxi\/]), Zhang et al.\ (1994; [\nevi\/], [\mgviii\/],
[\six\/]), Saraph \& Tully (1994; [\alv\/], [\sivi\/], [\sviii\/],
[\arx\/]), Butler \& Zeippen (1994; [\mgv\/], [\alvi\/], [\sivii\/],
[\suix\/], [\arxi\/]), Pelan \& Berrington (1995; [\fex\/]), Storey,
Mason \& Saraph (1996; [\fexiv\/]).  We discuss possible problems with
the new iron collision strengths in $\S$~3.3.  Most of the Einstein
transition probabilities were taken from Kaufman \& Sugar (1986), and
photoionization cross-sections are from Verner et al.\ (1996); many of
which are fits to cross-sections generated by the Opacity Project.  The
cross-sections of many of these ions have changed significantly from
those used by Korista \& Ferland (1989).  The other major uncertainty
in the results presented here, other than the iron coronal line
collision strengths, is the ionization balance of the third and fourth
row elements (most of the ions presented here), due to uncertainties in
the low-temperature dielectronic recombination rates for these elements
(see Nussbaumer \& Storey 1984 and Ali et~al.\ 1991).

\subsection{Assumptions}
We assume the origin of the coronal line emission to be from gas
primarily photoionized by the central AGN, though in some cases shocks
from strong radio jets may be important (Tadhunter et al.\ 1988; Morse,
Raymond, \& Wilson 1996).  The origin of the gas is presently unknown,
but may be related to galactic \hii\/ regions and molecular clouds
(Korista \& Ferland 1989; Oliva et al.\ 1994; Pier \& Voit 1995). We
have used the spectral synthesis code {\sc Cloudy} (version 90.02;
Ferland 1996) to calculate the emission from plane parallel, constant
hydrogen density clouds ionized by a continuum similar to a typical
Seyfert galaxy with $L_{ion} = 10^{43.5}$. The shape of the ionizing
continuum was chosen to be a combination of a UV-bump of the form
$f_{\nu} \propto \nu^{-0.3} exp(-h\nu /kT_{cut})$ and an X-ray power
law of the form $f_{\nu} \propto \nu^{-1.0}$ spanning 13.6~eV to
100~keV.  The UV-bump cutoff temperature, $T_{cut}$, was chosen such
that the UV-bump peaked (in $\nu F _{\nu}$) at 48~eV.  The UV and X-ray
components were combined with a typical Seyfert UV to X-ray spectral
slope, $\alpha_{ox} = -1.2$.

{\sc Cloudy} now considers the ionization balance of the first thirty
elements; we assume solar abundances from Grevesse \& Anders (1989) and
Grevesse \& Noels (1993).

\noindent
H :1.00E+00  He:1.00E-01  Li:2.04E-09  Be:2.63E-11  B :7.59E-10  
C :3.55E-04  N :9.33E-05  O :7.41E-04  F :3.02E-08  Ne:1.17E-04
Na:2.06E-06  Mg:3.80E-05  Al:2.95E-06  Si:3.55E-05  P :3.73E-07  
S :1.62E-05  Cl:1.88E-07  Ar:3.98E-06  K :1.35E-07  Ca:2.29E-06
Sc:1.58E-09  Ti:1.10E-07  V :1.05E-08  Cr:4.84E-07  Mn:3.42E-07
Fe:3.24E-05  Co:8.32E-08  Ni:1.76E-06  Cu:1.87E-08  Zn:4.52E-08

\noindent
While grains are present in at least the partially shielded portions of
narrow line region clouds (Ferland 1993), we will show ($\S$2.4) they
are likely absent in the coronal line emitting gas (see also Korista \&
Ferland 1989 and Oliva et al.\ 1994). 

The ionization/thermal equilibrium and radiative transfer calculations
for a single cloud proceeded until one of the following conditions was
met.  (1) The electron temperature dropped below 3000~K, (2) the
thickness of the cloud exceeded 10\% of its distance from the central
continuum source (in keeping with plane parallel clouds), or (3) the
total hydrogen column density exceeded 10$^{24}$ cm$^{-2}$.  In
practice the third condition had little impact on the coronal emission
lines.  The plane parallel condition was invoked to avoid clouds whose
dimensions rival their distances from the central source, thus avoiding
a significant covering fraction by a single cloud.  The largest cloud
of the grid presented below is one with a thickness of $\sim$\/16~pc,
consistent with the size of large molecular clouds.  We will comment
further on the effects of these stopping criteria in a later section.

Finally, we assume that each cloud sees the full continuum with no
obscuration.  At large enough distances, clouds or diffuse ISM may
attenuate the ionizing spectrum significantly.

\subsection{Reprocessing efficiency} 
Given the assumptions above, the emission from 1,881 individual clouds
has been computed as a function of the cloud distance from the ionizing
source, $\log R$ (cm), and the hydrogen number density of the cloud,
log $n(H)$ (cm$^{-3}$). We attempted to span all expected phase-space
in the cloud distance -- gas density plane necessary to emit the
coronal lines.  For the assumed source luminosity, the cloud distances
spanned approximately 0.5 light-days to 3~kpc; the gas density spanned
8 orders of magnitude 10$^2$ -- 10$^{10}$ cm$^{-3}$. Coronal line
emitting gas with significantly lower densities will have vanishingly
smaller surface brightnesses and thus will not be considered here. The
cloud distances will scale with luminosity as $L_{43.5}^{1/2}$, where
$L_{43.5}$ is the ionizing luminosity in units of 10$^{43.5}$ ergs
s$^{-1}$.  Note that this scaling with observed luminosity assumes that
the continuum is emitted isotropically, which may not be the case in
nature.  Recent observations have shown emission line ``cones'' (Evans
et al.\ 1991; Tsvetanov \& Walsh 1992; Wilson et al.\ 1993; Macchetto
et al.\ 1994; and Arribas et al.\ 1996) and UV photon-deficits
(Binette, Fosbury, \& Parker 1993; and Morse, Raymond \& Wilson 1996).
This ``beaming'' of the incident continuum, which the clouds see and we
may not, will change the effective luminosity incident on the clouds
and adjust the distances presented below.

Contour plots of emission line equivalent width, referred to the
incident continuum at 4860~\AA\/, as a function of $\log R$ and $\log
n(H)$ for 24 high ionization forbidden lines are shown in
Figures~1--4.  This equivalent width assumes 100\% source coverage by
the emitting clouds. We show those lines that are either expected to be
strong and/or observationally important. The emission line equivalent
width distributions are plotted in order of increasing atomic number,
then increasing ionization stage for each element, and indicate the
efficiency of continuum reprocessing.

For each emission line in Figures~1--4, a ridge of near maximum
equivalent width runs diagonally across the distance -- gas density
plane, roughly parallel to lines of constant ionization parameter. The
ionization parameter, $U(H)$, is defined as the ratio of ionizing
photon density to hydrogen density, $U(H) \equiv \Phi(H)/n(H)c$, where
$\Phi(H)$ is the flux of ionizing photons and $c$ is the speed of
light. This parameter increases from top right to bottom left in these
diagrams, and $\log U(H) = -4.0$ is plotted as a dashed line for each
emission line as a reference.  This ridge of near maximum equivalent
width coincides with a small range in optimal ionization parameter. For
larger (smaller) ionization parameters the gas is over (under) ionized,
and the line is not efficiently emitted. This explains the sudden drops
in the emission line equivalent widths on either side of their ridges.
The asymmetric drop of emission on either side of the ridges is
explained by the differing physical conditions of the gas on either
side of the ridges and by our assumption of plane parallel clouds
($\S$2.2).  The latter has the effect of truncating the emission of
some of the ``lower ionization'' coronal lines, such as [\nev\/], at
high ionization parameter where they would otherwise form on the back
sides of the clouds.  In the absence of this truncation, the ridges of
optimal emission would broaden slightly in the direction of higher
ionization parameter. The effect on the higher ionization coronal lines
is minimal since these lines generally form in a thin zone on the side
of the cloud facing the ionizing source.

Moving {\em along} the ridge to increasing gas densities, at near
constant ionization parameter, the forbidden lines become collisionally
deexcited and the line equivalent width falls off.  Compare, for
example, in Figures~1a,b, the near-UV and infrared transitions of
[\nev\/].  Moving along the ridge at constant ionization parameter to
smaller gas densities, the equivalent widths of some lines also
diminish as other lines of similar ionization, but lower critical
density, become important coolants.  These contour plots represent
``visibility functions'' for the coronal lines.

In Table~1 we give the ionization parameter at the peak equivalent
width (column~3), the range in $\log n(H)$ for the top contour
(column~4) of each line considered in Figures~1 -- 4. Together these
parameters describe the optimal conditions in which these lines form.
Note that a range in gas density for a given ionization parameter
indicates a range of expected distances from the central source for
$L_{ion} = 10^{43.5}$ ergs~s$^{-1}$, given in column~5 of Table~1. We
emphasize that these distance ranges are not {\em necessarily} those
for which the coronal lines should be most luminous in Seyfert
galaxies. For example, we do not expect a substantial contribution of
coronal line emission from the broad emission line region. Rather,
column~5 is meant to illustrate the relative differences in the coronal
line formation distances. In column~6 we give the $\log T_e$ (K) of the
front face of the cloud whose ionization parameter is given in
column~3. This temperature is representative of that in which the line
in column~1 is emitted. Column~7 lists the peak equivalent width of
each line, as indicated in Figure~1.  Since this equivalent width is
referenced to the same point in the incident continuum, a comparison
should grossly indicate the expected relative strengths of these lines,
though this is not meant to represent a predicted coronal emission line
spectrum that is outside the scope of the present paper.

\subsection{Dust in the coronal line gas?}
Pier \& Voit (1995) hypothesized that the coronal line emission comes
from a thin, highly ionized ``skin'' just above the surface of the
``obscuring'' molecular torus which is undergoing evaporation from an
X-ray heated wind generated by the central continuum source. In this
model, the grains are not destroyed until they are sputtered in the
$10^6$~K wind, with some destruction possible in the coronal line
emitting gas just below the wind. However, their predicted coronal line
spectrum assumed solar abundances and did not take into account gas
phase depletions onto grains (Pier 1995). In this section we
demonstrate the full effects of the presence of grains in a more
general coronal line emitting environment.

There are three major effects of dust on line formation: (1) the
emission lines weaken due to absorption of the incident continuum by
dust at large $U(H)$, (2) the grains photoelectrically heat the gas,
and (3) some of the gas-phase elements are depleted. The destruction of
the coronal line photons by grains is included but not important since
the line and IR continuum optical depths are small, in comparison to
the resonance lines which are readily destroyed by grains. We include
dust of the type found in the Orion nebula (Baldwin et al.\ 1991) in
grid calculations at radii consistent with the sublimation temperatures
of the silicate and graphite grains discussed by Laor \& Draine (1993)
and Netzer \& Laor (1993). In particular, inside the graphite grain
boundary the gas had solar abundances and no grains were present. The
Orion type graphite grains were turned on at a radius of $10^{16.9}$~cm
(where T$_{dust}$(graphite) $\approx$ 1750~K) and the carbon abundance
was set to the value given below. The Orion type silicate grains were
turned on at a radius of $10^{17.6}$~cm (where T$_{dust}$(silicate)
$\approx$ 1400~K), and the gas abundances of the important elements
then took on those of the ionized gas in the Orion nebula (see Baldwin
et al.\ 1996):

\noindent
 H :1.00E+00  He:9.50E-02  C :3.00E-04  N :7.00E-05  O :4.00E-04  
 Ne:6.00E-05  Na:3.00E-07  Mg:3.00E-06  Al:2.00E-07  Si:4.00E-06  
 S :1.00E-05  Cl:1.00E-07  Ar:3.00E-06  Ca:2.00E-08  Fe:3.00E-06  
 Ni:1.00E-07  

\noindent
These abundances are based upon the results of several recent studies
of the Orion nebula (Baldwin et~al.\ 1991, Rubin et~al.\ 1991, 1992a,b,
and Osterbrock, Tran \& Veilleux 1992,). Note that while nitrogen and
neon are not expected to be depleted onto grains, the current best
estimates of their gas phase abundances are found to be 75\% and 50\%
of their solar values relative to hydrogen, respectively.  O/H and C/H
are also depressed, their abundances being 54\% and 85\% solar,
respectively. How much of these differences from solar are due to
depletions onto grains or due to intrinsic abundance differences is
unknown and is an area of active research (e.g., Snow \& Witt 1996).
Since carbon and oxygen do much of the line cooling, some differences
in the cloud thermal structure and emission will result, independent of
the effects of the grains themselves.  A full treatment of the dust
physics was included in the simulations (see Baldwin et al.\ 1991).

In Figure~5 we show the equivalent width contours of six coronal lines
formed in the presence of dust, as described above. These lines should
be compared to their solar abundance, dust-free gas counterparts in
Figures~1 -- 4. A comparison of Figure~5f with 4f ([\fexiv\/]
$\lambda$5303) illustrates most dramatically the effects on the line
emission. At radii smaller than $10^{16.9}$~cm, grains are not present
and the contours shown in Figure~5f are exactly those shown in
Figure~4f.  For radii just outside this graphite grain sublimation
boundary, iron is not depleted, but a significant decline in the line
equivalent width (factor of $\sim 6$) occurs due to the first effect,
mentioned above.  This line's equivalent width again falls rapidly just
beyond the silicate grain sublimation radius (another factor of $\sim
25$). At these larger radii many of the elements, including iron, are
depleted onto grains.  Some of the more refractory elements and their
Orion depletion factors are Mg(12), Si(12), Ca(115), Fe(11). Note that
these depletions are less severe than in the local ISM, and could in
part represent grain destruction in the ionized gas (Rubin et
al.\ 1992a,b).  Mainly because of its extreme depletion in a dusty
environment, the emission lines of Ca are devastated.  (Figure~5e). The
peak equivalent width contours of [\mgviii\/] 3.03$\mu$m,
[\six\/] 1.43$\mu$m, and [\caviii\/] 2.32$\mu$m in Figure~5
have smaller values and have moved to smaller radii along their ridges,
where the effects of grains are smaller or absent.  The effects of the
presence of grains on the emission of [\nev\/] $\lambda$3426 and
[\nevi\/] 7.65$\mu$m are more subtle.  Neon, a noble gas, is not
depleted and these lines are emitted almost entirely outside the grain
sublimation radii.  These lines suffered declines in their peak
equivalent widths of factors of 2 and 3, respectively, and their ridges
shifted to lower ionization parameters where the grain/gas opacity
ratio is smaller.  This effect is not as important for these two lines,
compared to the others in Figure~5, since they form at smaller
ionization parameters.

Figures~1 -- 5 and column 5 of Table 1 show that the lower ionization
coronal lines form in regions outside the dust sublimation radii, while
the higher ionization coronal lines form in regions which straddle
these radii. Given the severe reduction in many of the coronal line
intensities in a dusty environment, the observed strengths of the
coronal emission lines indicate that they are formed in dust-free, or
nearly so, gas. The observations in several Seyfert galaxies of
significant [\caviii\/] 2.32$\mu$m emission relative to [\nev\/]
$\lambda$3426 and [\sivii\/] 2.48$\mu$m, all formed mainly outside the
sublimation radii (Oliva et al.\ 1994; Marconi et al.\ 1996) would by
itself seem to exclude the possibility of the existence of dust within
the coronal line gas. We elaborate further on this, below.

\section{Discussion}

\subsection{Coronal Line Formation in Photoionized Gas} 
The coronal lines shown in Figure~1 form in gas that has typical
ionization parameters $-2.0 \ltsim \log U(H) \ltsim 0.75$ and gas
densities $2.0 \ltsim \log n(H) \ltsim 8.5$ (cm$^{-3}$) and
temperatures $\sim$ 12,000 K -- 150,000 K.  This corresponds to
distances from the ionizing source of approximately the broad line
region to 400~pc ($L_{43.5}^{1/2}$).  This upper limit on the distance
could be larger if lower density gas is present, but this gas would
have a smaller surface brightness. It is interesting to note that the
lower ionization coronal lines, [\nev\/] to [\sivii\/] and
also [\caviii\/] and [\fevii\/] form most efficiently in gas
that is roughly $\sim$10~pc and beyond, and thus their emission may be
extended in nearby Seyfert galaxies (Korista \& Ferland 1989).  The
rest of the lines in Table~1 optimally form in gas less than 10~pc from
the ionizing source.  Looking at Table~1 we expect that [\fevii\/]
$\lambda$6087 and [\fexiv] $\lambda$5303 form in completely
different gas. For comparison [\oiii] $\lambda$5007 is expected
to form in gas at distances of several parsecs to $\sim 1$~kpc.

If there is a relatively simple relation between the cloud distance and
its velocity, then one might expect to find observationally a
relationship between the width of the line and its ionization potential
and/or critical density; the latter two are related. Such relationships
have been observed amongst the optical forbidden lines (e.g.,
Filippenko \& Halpern 1984; DeRobertis \& Osterbrock 1984, 1986).
Based on Figures~1a and 1b and columns 3 -- 5 in Table~1, one might
expect that the infrared [\nev\/] 14.3,24.3~$\mu$m lines would
be primarily emitted in gas of very different properties than that of
the [\nev\/] $\lambda$$\lambda$3426,3346 lines, {\em even though
they arise from the same ion}. The near-UV line profiles are likely to
be significantly broader than their infrared counterparts. To our
knowledge the infrared lines of [\nev\/] have not been reported
for Seyfert galaxies, but should be observed with ISO.

The equivalent widths of the coronal lines should serve as an effective
Zanstra temperature (Zanstra 1931; Osterbrock 1989) of the relevant
ionizing continuum ($\sim 0.1 - 0.5$~keV).  In $\S$2.2 we assumed a
fairly hard continuum incident on the coronal line clouds. Decreasing
the coronal ionizing photon flux, $\Phi$(0.1 -- 0.5~keV), by a factor
of 10 will decrease the equivalent width of the higher ionization
lines, such as [\fex\/] and [\six\/], by factors of 10
and 20, and the lower ionization lines, such as [\nev\/], by
factors of 2 -- 3.  A more subtle effect of the shape of the ionizing
spectrum is that the formation distances of especially the highest
ionization coronal lines (column~5 of Table~1) will shift to smaller
(larger) values for significantly softer (harder) spectra, due to
changes in the thermal/ionization balance in the clouds. In the example
of the diminished $\Phi$(0.1 -- 0.5~keV), above, the formation
distances in column~5 of Table~1 would be $\sim$~2 times smaller for
the highest ionization lines (e.g., [\fex\/] $\lambda$6375 and
[\fexiv\/] $\lambda$5303), while the effects on lines such as
[\nev\/] and [\fevii\/] would be much smaller. Many of the
highest ionization lines might be invisible, too.

\subsection{Relative abundance determinations}
Nussbaumer \& Osterbrock (1970) suggested using the [\fevii\/]
$\lambda$6087 / [\nev\/] $\lambda$3426 ratio to measure the
Fe/Ne abundance. In principle this would be a good ratio, if the two
lines are formed in the same gas, since neon is not expected to be
depleted when grains are present. However, a glance at Table~1 and
Figures~1a and 4c shows that these lines generally do not form in the
same gas. In this section we discuss a few coronal line ratios which
should prove to be reliable gas abundance indicators.

The ratios of emission lines forming in gas with the same physical
conditions, i.e. gas density, temperature and ionization, will be good
relative abundance indicators.  Considering Figures~1 -- 4 and Table~1,
we have identified several pairs of lines that form in similar physical
conditions.  Their ridges of peak equivalent widths overlie one
another, and their ratio should be nearly flat in the gas density --
source distance plane.  Figure~6 shows the best example of line
ratios that overlap well in the gas density -- source distance plane.
The ratio [\caviii\/] 2.32$\mu$m/[\sivii\/] 2.48$\mu$m
has a nearly flat plateau, with a linear value between 1.4 and 1.5
for solar gas abundances, though values as small as 1 are possible in
some gas where the two lines are strong.  Table~2 summarizes the eight
most reliable abundance indicators from considering Table~1.  Column
(2) of the table gives the value of the plateau of the line ratio.  For
two of the ratios, [\alvi\/]/[\sivi\/], and
[\arx\/]/[\fexi\/], the plateau was fairly flat due to
the ridges of the individual lines overlapping well, with errors of
$\sim$~30\% -- 40\%.  However, the other ratios were not as well
behaved and the plateau value is only known to a factor of two
indicating the sensitivity to the line formation conditions.  The
observed line ratios, if any, are given in column (3) for two Seyfert
galaxies, and the solar abundance ratio is in column (4).

From Table~2 we see that the ratio of [\caviii\/]/[\sivii\/]
is predicted, with solar abundances, to be 2--3 times that of the
observed value.  Taking the observed ratios at face value, either
calcium is underabundant relative to silicon by a factor of two, or the
reverse, silicon is overabundant relative to calcium compared to
solar.  This is nowhere near the factor of 13 underabundance of calcium
relative to silicon in the Orion nebula (Rubin et al.\ 1992a,b), or the
factor of 225 in the local ISM (Snow \& Witt 1996), supporting our
earlier statement ($\S$~2.4) that the coronal line emitting gas is dust
free.  This has important implications for the origin of the coronal
line gas, as well as the ``classical'' narrow emission line gas since
there must be some spatial overlap. Are the narrow line region clouds
dusty throughout? Does the sublimation of dust demarcate the narrow and
broad emission line regions (Netzer \& Laor 1993)? If the source of gas
for the coronal lines is dusty (e.g., molecular clouds or torus), how
are the grains destroyed in the emitting region? The electron
temperatures at the front faces of the coronal line emitting clouds
range from $\sim$~12,000~K -- 150,000~K, and the destruction of grains
due to sputtering is probably unimportant, except on long time scales
(Draine \& Salpeter 1979). However, some mechanism must be responsible
for the destruction of most of the dust in the coronal line emitting
gas.

\subsection{The iron conundrum}
Oliva et al.\ (1996) highlighted the sensitivity of the collision
strengths to the presence and position of resonances in the collisional
cross sections.  The collision strengths of [\fex\/] and
[\fexiv\/] are now more than an order of magnitude larger than
those used in Korista \& Ferland (1989) which did not include the
effects of resonances (Mason 1975). However, even some of the recent
calculations which include resonances differ significantly in their
resulting collision strengths. For example, the most recently computed
collision strengths of [\fex\/] and [\fexiv\/] have
differed by factors of 3 -- 10 (Mohan, Hibbert, \& Kingston 1994; Pelan
\& Berrington 1995; Dufton \& Kingston 1991; Storey, Mason, \& Saraph
1996). No collision strengths which include resonances have been
calculated for [\fexi\/], and those which do for [\fevii\/]
(Keenan \& Norrington 1987) may be in doubt.  Preliminary analysis (see
also column~7 of Table~1) indicates that if the most recent collision
strengths for [\fex\/] and [\fexiv\/] are correct (Pelan
\& Berrington; Storey et al.), then the iron abundance would have to be
arbitrarily depleted relative to the other elements (e.g., Mg, Si, Ca)
by more than a factor of $\sim 10$ compared to solar, which does not
make sense.  The alternative is that the [\fex\/] and
[\fexiv\/] collision strengths are too strong by a similar
factor.  Hopefully, these issues will be resolved in the future.  The
line ratios involving Fe are currently uncertain, but these and the
relative abundance ratios should scale roughly with changes in the iron
lines' collision strengths.

\section{Summary}
Using a large number of photoionization simulations we have illustrated
the physical conditions (gas density and distance from the ionizing
source) in which the coronal emission lines in Seyfert galaxy spectra
form. We find that the observations of significant emission from ions
of refractive elements such as calcium and iron likely excludes the
presence of dust grains in the coronal line emitting gas. The lower
ionization coronal lines (e.g., [\nev\/], [\fevii\/]) are
likely to form in lower density gas which should be spatially extended
in nearby Seyfert galaxies, whereas the highest ionization lines
(e.g.,[\six\/], [\fexiv\/]) are likely to form in higher
density gas in more compact regions.  We expect that the coronal line
region to be up to $\sim 400L_{43.5}^{1/2}$ pc in size.  Spatially
resolved spectroscopy of the optical and infrared coronal lines would
prove invaluable to our understanding of these high ionization emission
lines.

\acknowledgments
We thank NASA and NSF for support through NAG-3223 and AST93-19034, and
STSCI for GO-06006.02.

%
%
%
%
\newpage
\begin{center}
{\bf Figure Captions}
\end{center}

\noindent
Fig.~1.--- {Contours of constant logarithmic line equivalent widths as
a function of $\log R$ and $\log n(H)$ for the ions indicated,
referenced to the incident continuum at 4860\AA\/.  The bold lines
represent 1 dex and the dotted lines are 0.2 dex steps.  The triangle
is the peak of the equivalent width distribution and the contours
decrease downward to the outer value of 1~\AA\/. The reader will
sometimes find it convenient to view the contour plots {\em along} the
ridge at large inclination angle to the sheet of paper.}

\noindent
Fig.~2.--- {Same as Fig.~1 for the ions indicated.}

\noindent
Fig.~3.--- {Same as Fig.~1 for the ions indicated.}

\noindent
Fig.~4.--- {Same as Fig.~1 for the ions indicated.}

\noindent
Fig.~5.--- {Same as Fig.~1 for the ions indicated including dust
as described in $\S$~2.4.}

\noindent
Fig.~6.--- {Logarithmic contours of the ratio [\caviii\/]
2.32$\mu$m/[\sivii\/] 2.48$\mu$m.  The bold outer contour is a linear
ratio of 1, and the dotted lines are 0.1 dex steps.  The linear value
of the plateau ratio is between 1.4 and 1.5.}

\newpage
%
\begin{table}[h]
\begin{center}
\begin{tabular}{ccccccc}
\multicolumn{7}{c}{\sc TABLE~1} 
\\[0.1cm]
\multicolumn{7}{c}{\sc Coronal Line Physical Conditions}
\\[0.2cm]
\hline
\hline
%
\multicolumn{1}{c}{Ion}
&\multicolumn{1}{c}{$\lambda$$^a$}
&\multicolumn{1}{c}{$\log U(H)^b$}
&\multicolumn{1}{c}{range $\log n(H)$$^c$} 
&\multicolumn{1}{c}{range $\log R^d$}
&\multicolumn{1}{c}{peak log T$_e$$^e$}
&\multicolumn{1}{c}{$W_{\lambda4860}^{max}$$^f$}\\
\multicolumn{1}{c}{(1)} & \multicolumn{1}{c}{(2)} & 
\multicolumn{1}{c}{(3)} & \multicolumn{1}{c}{(4)} & 
\multicolumn{1}{c}{(5)} & \multicolumn{1}{c}{(6)}
& \multicolumn{1}{c}{(7)}
\\[0.05cm]
\hline
$[$\nev\/] & $\lambda$3426 & $-$1.25 & L --  7.0 & 18.1 -- 20.6 & 4.3 & 621. \\
$[$\nev\/] & 14.3$\mu$m & $-$1.5 & L --  4.0 & 19.8 -- 20.8 & 4.2 & 930. \\
$[$\nevi\/] & 7.65$\mu$m & $-$0.75 & 3.0 -- 4.0 & 19.4 -- 19.9 & 4.4 & 1728. \\
$[$\navi\/] & 14.3$\mu$m & $-$1.5 & L -- 4.0 & 19.8 -- 20.8 & 4.2 & 11.6 \\
 
$[$\mgv\/] & 5.60$\mu$m & $-$2.0 & L -- 6.0 & 19.0 -- 21.0 & 4.2 & 63.2 \\
$[$\mgvii\/] & 5.50$\mu$m & $-$1.0 & 2.5 -- 5.5 & 18.8 -- 20.3 & 4.3 & 184. \\
$[$\mgviii\/] & 3.03$\mu$m & $-$0.5 & 3.5 -- 6.0 & 18.3 -- 19.5 & 4.4 & 298. \\

$[$\alvi\/] & 3.66$\mu$m & $-$1.75 & L -- 5.5 & 19.1 -- 20.9 & 4.2 & 48.3 \\
$[$\alviii\/] & 5.85$\mu$m & $-$0.5 & 2.75 -- 5.75 & 18.4 -- 19.9 & 4.4 & 12.8 \\

$[$\sivi\/] & 1.96$\mu$m & $-$1.75 & L -- 7.75 & 18.0 -- 20.9 & 4.2 & 93.1 \\
$[$\sivii\/] & 2.48$\mu$m & $-$1.0 & 2.25 -- 6.75 & 18.1 -- 20.4 & 4.3 & 110. \\
$[$\siix\/] & 3.94$\mu$m & $-$0.25 & 3.0 -- 6.25 & 18.0 -- 19.6 & 4.5 & 256. \\
$[$\six\/] & 1.43$\mu$m & 0.25 & 5.25 -- 7.0 & 17.4 -- 18.3 & 4.8 & 284.

\\[0.01cm]
\hline
\end{tabular}
\end{center}
\end{table}

\clearpage

\begin{table}[h]
\begin{center}
\begin{tabular}{ccccccc}
\multicolumn{7}{c}{\sc TABLE~1 Continued} 
\\[0.1cm]
\multicolumn{7}{c}{\sc Coronal Line Physical Conditions}
\\[0.2cm]
\hline
\hline
%
\multicolumn{1}{c}{Ion}
&\multicolumn{1}{c}{$\lambda$$^a$}
&\multicolumn{1}{c}{$\log U(H)^b$}
&\multicolumn{1}{c}{range $\log n(H)$$^c$} 
&\multicolumn{1}{c}{range $\log R^d$}
&\multicolumn{1}{c}{peak log T$_e$$^e$}
&\multicolumn{1}{c}{$W_{\lambda4860}^{max}$$^f$}\\
\multicolumn{1}{c}{(1)} & \multicolumn{1}{c}{(2)} & 
\multicolumn{1}{c}{(3)} & \multicolumn{1}{c}{(4)} & 
\multicolumn{1}{c}{(5)} & \multicolumn{1}{c}{(6)}
& \multicolumn{1}{c}{(7)}
\\[0.05cm]
\hline

$[$\sviii\/] & $\lambda$9915 & $-$0.75 & 3.5 -- 8.25 & 17.3 -- 19.6 & 4.5 & 28.5 \\
$[$\suix\/] & 1.25$\mu$m & $-$0.5 & 3.0 -- 7.75 &  17.4 -- 19.8 & 5.0 & 143. \\
$[$\suxi\/] & 1.92$\mu$m & 0.5 & 4.75 -- 7.5 & 17.0 -- 18.4 & 5.1 & 20.2 \\

$[$\arvi\/] & 4.53$\mu$m & $-$1.75 & L -- 4.25 & 19.8 -- 20.9 & 4.2 & 166. \\
$[$\arx\/] & $\lambda$5533 & 0.0 & 5.0 -- 8.5 & 16.8 -- 18.5 & 4.6 & 11.6 \\
$[$\caviii\/] & 2.32$\mu$m & $-$1.25 & L -- 6.25 & 18.5 -- 20.6 & 4.3 & 152. \\

$[$\fevii\/] & $\lambda$6087 & $-$2.0 & 3.0 -- 7.5 & 18.3 -- 20.5 & 4.2 & 21.2 \\
$[$\fevii\/] & 9.51$\mu$m & $-$2.0 & L -- 5.0 & 19.5 -- 21.0 & 4.2 & 71.7 \\
$[$\fex\/] & $\lambda$6375 & $-$0.5 & 3.0 -- 8.25 & 17.1 -- 19.8 & 4.6 & 968. \\
$[$\fexi\/] & $\lambda$7892 & 0.0 & 5.0 -- 6.5 & 17.8 -- 18.5 & 4.6 & 41.2 \\
$[$\fexiv\/] & $\lambda$5303 & 0.75 & 5.75 -- 8.0 & 16.6 -- 17.8 & 5.2 & 468.

\\[0.01cm]
\hline
\end{tabular}
\end{center}
\rm
\footnotesize
\leftskip=2.em
$^a$Wavelengths in \AA\/ for the optical lines or $\mu$m for the IR lines.\\
\leftskip=2.em
$^b$Value of the dimensionless ionization parameter at the triangle in Fig. 1.\\
\leftskip=2.em
$^c$Minimum and maximum $\log n(H)$ (cm$^{-3}$).  Values of L indicates
that \\ the top ridge extends below the cutoff value of log $n(H) = 2$.\\
\leftskip=2.em
$^d$Minimum and maximum $\log R$ (cm) corresponding to the maximum \\
and minimum $\log n(H)$ for the peak $\log U(H)$ in column~3.\\
\leftskip=2.em
$^e$Log electron temperature (K) at front face of the cloud with $\log
U(H)$ \\ given in column 3.\\
\leftskip=2.em
$^f$Peak $W_{\lambda4860}$ (\AA\/) of line in Figure~1 for a cloud
covering factor of 1.
\end{table}
\newpage
\begin{table}[h]
\begin{center}
\begin{tabular}{cccc}
\multicolumn{4}{c}{\sc TABLE~2} 
\\[0.1cm]
\multicolumn{4}{c}{\sc Abundance Ratio Indicators}
\\[0.2cm]
\hline
\hline
%
\multicolumn{1}{c}{Line Ratio$^a$}
&\multicolumn{1}{c}{Plateau Ratio$^b$}
&\multicolumn{1}{c}{Observed Ratios}
&\multicolumn{1}{c}{Ratio$_{\odot}^c$}\\
\multicolumn{1}{c}{(1)} & \multicolumn{1}{c}{(2)} & 
\multicolumn{1}{c}{(3)} & \multicolumn{1}{c}{(4)} 
\\[0.05cm]
\hline

$[$\alvi\/]/$[$\sivi\/] & 0.6  &  --  & 0.083\\
$[$\arvi\/]/$[$\sivi\/] & 1.8 -- 2.2  &  -- &  0.11\\

$[$\mgvii\/]/$[$\sivii\/] & 1.3 -- 2.5  &  --  & 1.07\\
$[$\caviii\/]/$[$\sivii\/] & 1.4 -- 1.5  &  0.81$^d$, 0.46$^e$ &  0.065\\

$[$\mgviii\/]/$[$\six\/]  & 1.8 -- 3.0  & -- & 2.4 \\
$[$\alviii\/]/$[$\suix\/]  & 0.07 -- 0.14  & -- & 0.18 \\

$[$\mgv\/]/$[$\fevii\/]$^f$ & 0.7 -- 0.9  &  --  & 1.17\\
$[$\arx\/]/$[$\fexi\/] & 0.26  &  --  & 0.123
\\[0.01cm]
\hline
\end{tabular}
\end{center}
\rm
\footnotesize
\leftskip=2.em
$^a$ratio of overlapping lines in Table 1.\\
\leftskip=2.em
$^b$predicted line ratio where nearly constant, or a range \\
is given if
the equivalent widths of the two lines do not \\ closely overlap.\\
\leftskip=2.em
$^c$solar abundance ratio of the elements in column (1)\\
\leftskip=2.em
$^d$NGC~1068, Marconi et al.\ 1996\\
\leftskip=2.em
$^e$Circinus galaxy (A1409$-$65), Oliva et al.\ 1994\\
\leftskip=2.em
$^f$$[$\fevii\/] 9.51$\mu$m 
\end{table}
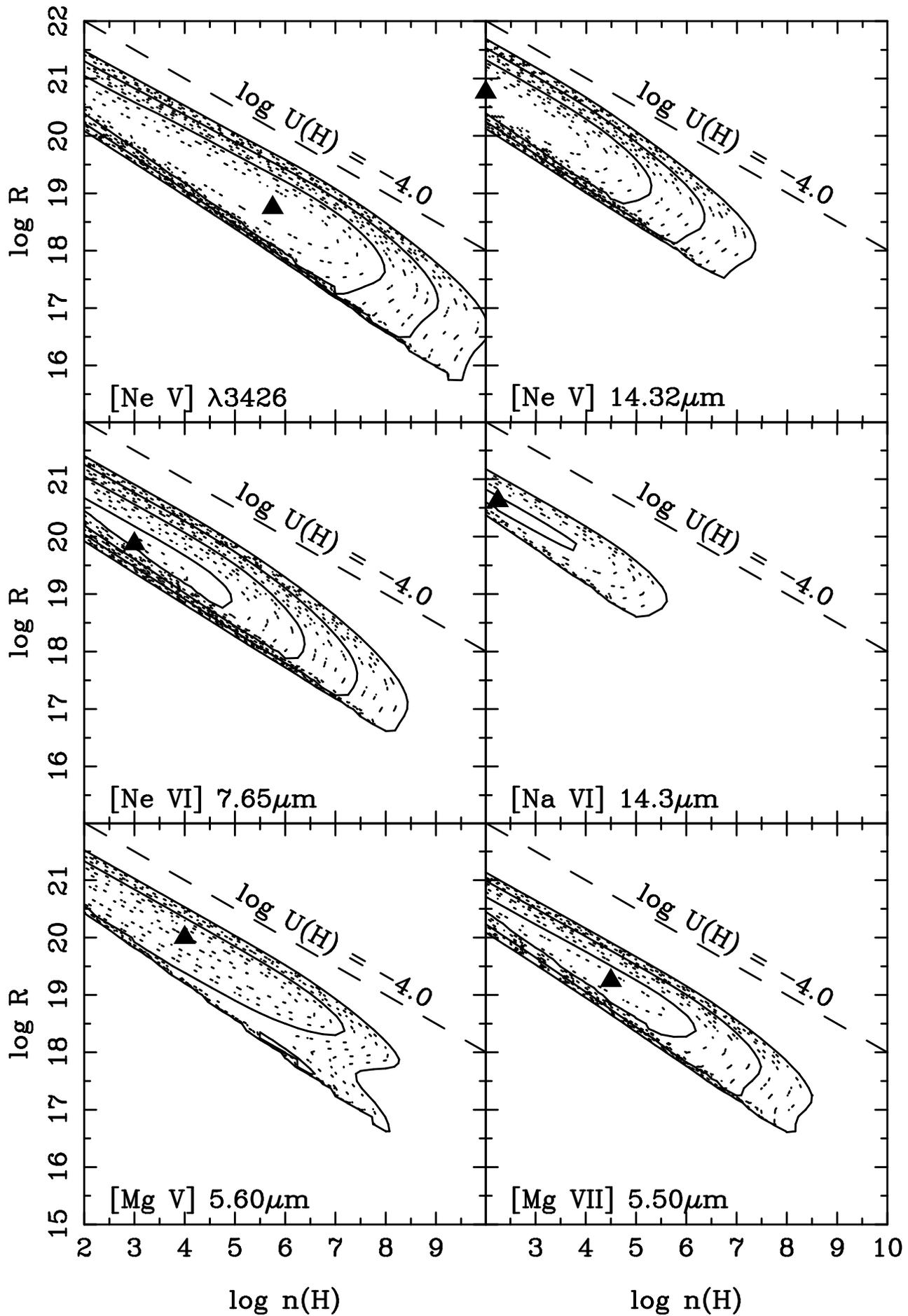
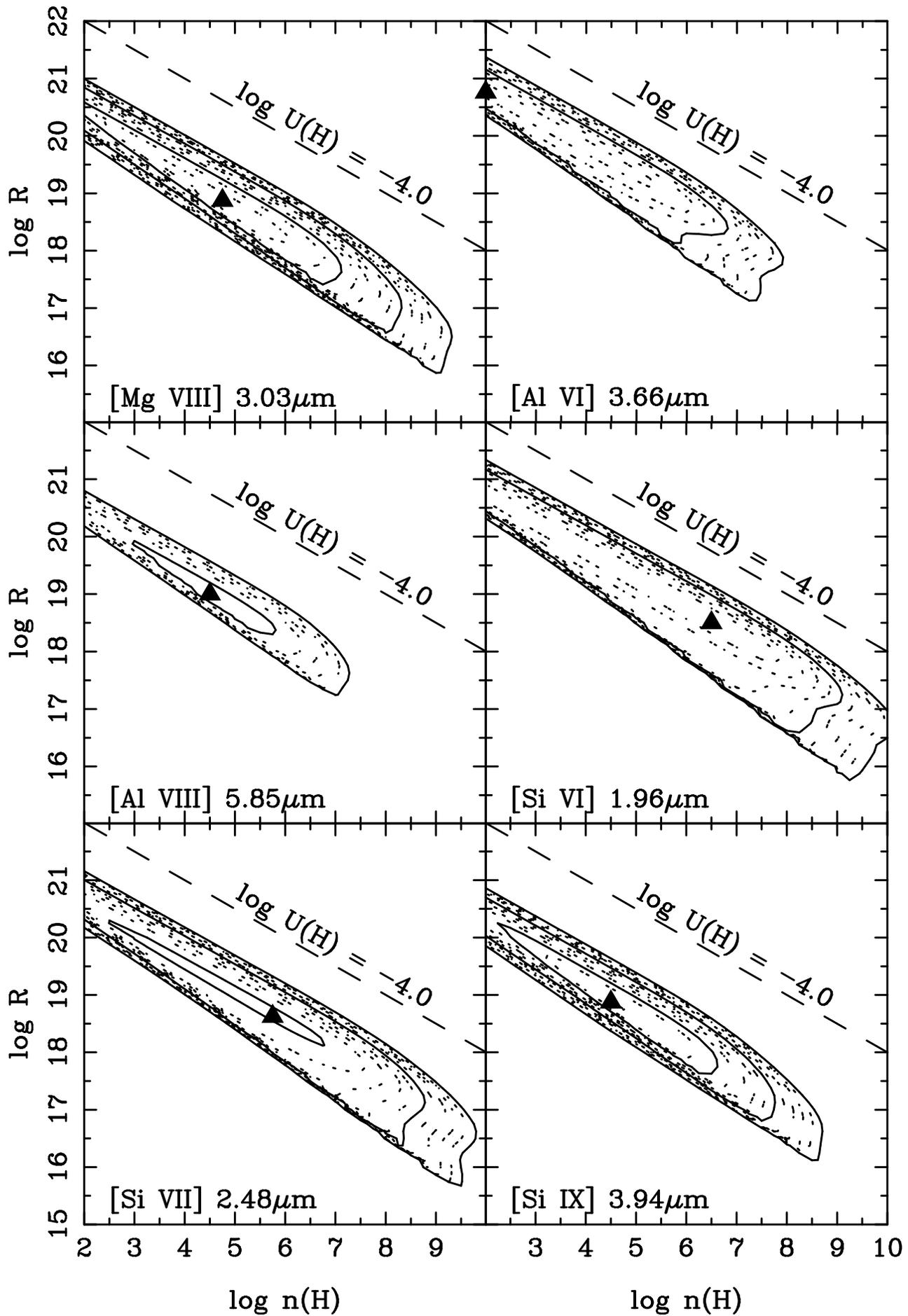
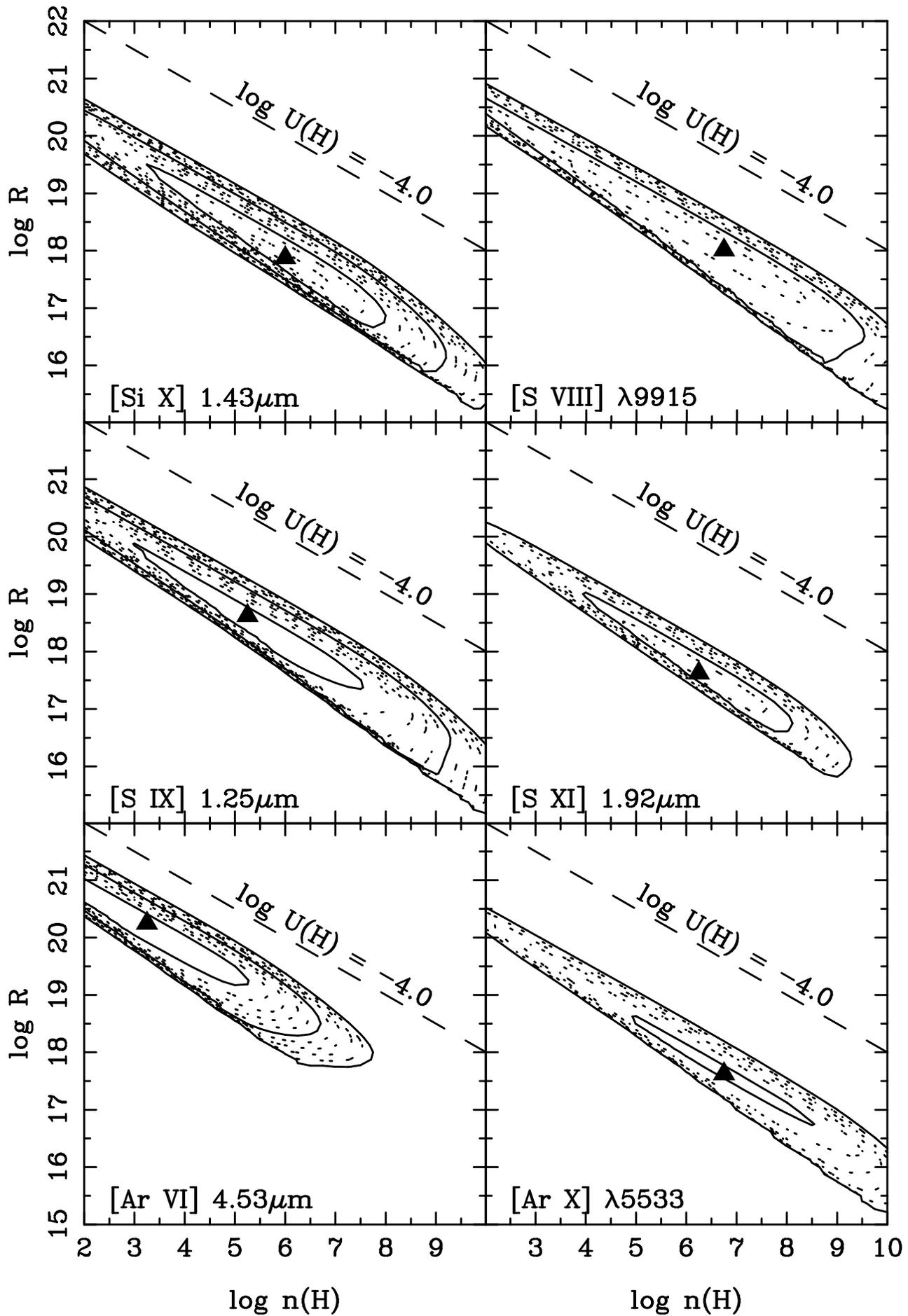
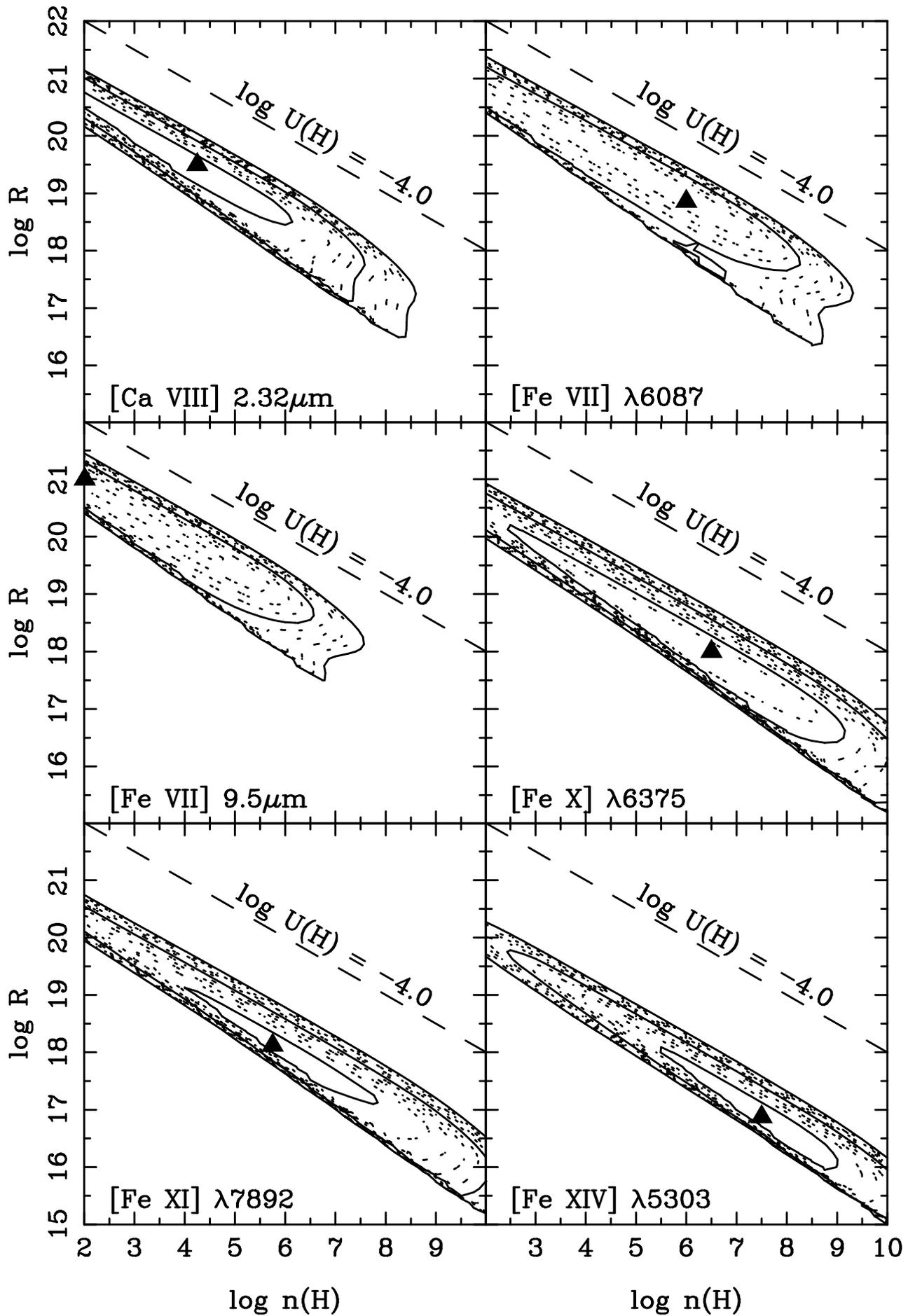
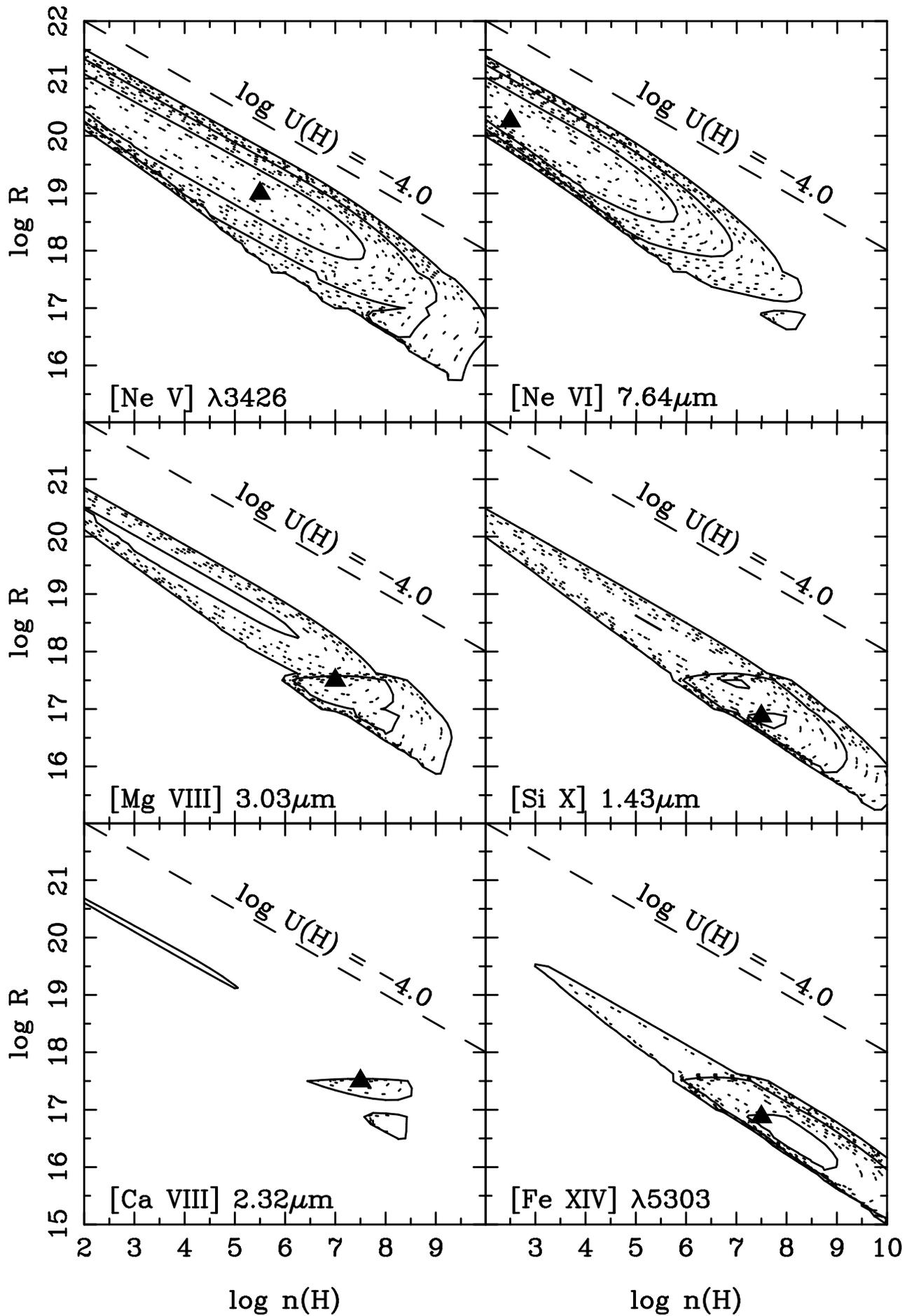
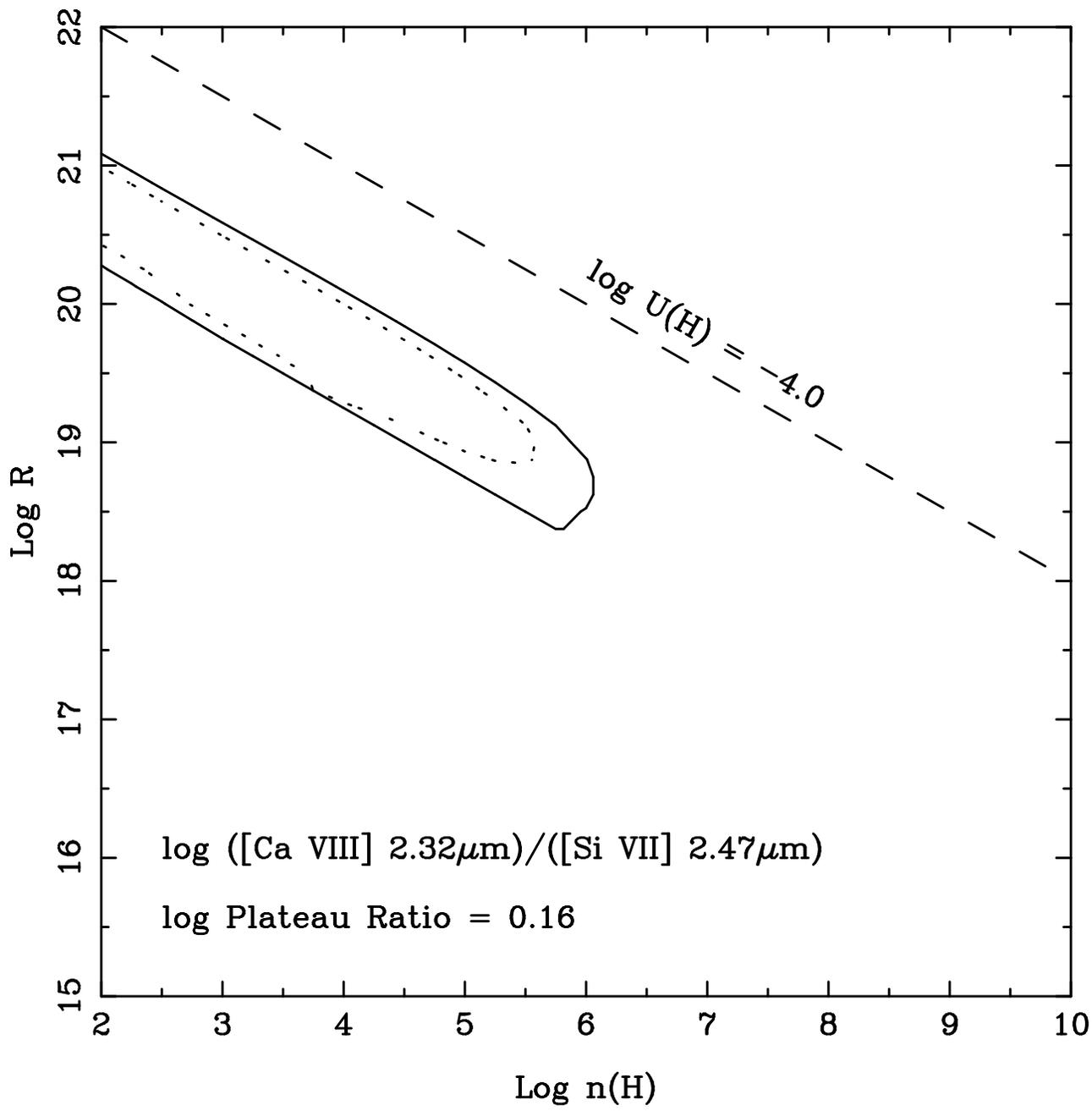
\end{document}